\documentclass[a4paper]{article}
\usepackage{amscd,amsmath,amssymb}
\usepackage[usenames, dvipsnames]{color}

\newcommand{\p}[1]{(\ref{#1})}

\newcommand{\cA}{{\cal A}}

\newcommand{\cV}{{\cal V}}

\newcommand{\tW}{\widetilde{W}}
\newcommand{\tU}{\widetilde{U}}

\newcommand{\bQ}{{\overline Q}{}}

\newcommand{\bpsi}{{\bar\psi}{}}

\newcommand{\bu}{{\bar u}}

\newcommand{\sfrac}[2]{{\textstyle\frac{#1}{#2}}}
\newcommand{\und}{\qquad\textrm{and}\qquad}
\renewcommand{\=}{\ =\ }

\newcommand{\be}{\begin{equation}}
\newcommand{\ee}{\end{equation}}
\newcommand{\bea}{\begin{eqnarray}}
\newcommand{\eea}{\end{eqnarray}}

\newcommand{\ba}{\begin{array}} \newcommand{\ea}{\end{array}}

\def\im{{\rm i}}

\newcommand{\nn}{\nonumber}

\begin{document}

\pagenumbering{gobble}

\begin{flushright}
%\today
\end{flushright}

\vspace{2cm}

\begin{center}
{\LARGE\bf ${\cal N}{=}\,4$ supersymmetric mechanics on curved spaces}
\end{center}

\vspace{1cm}

\begin{center}

{\Large Nikolay~Kozyrev${}^{a}$, Sergey~Krivonos${}^{a}$ , Olaf~Lechtenfeld${}^{b}$, \\[8pt]
  Armen~Nersessian${}^{a,c}$ and Anton~Sutulin${}^{a}$
}
\end{center}

\vspace{5mm}

\begin{center}
${}^a$ {\it
Bogoliubov  Laboratory of Theoretical Physics, JINR,
141980 Dubna, Russia}

\vspace{2mm}

${}^b$ {\it
Institut f\"ur Theoretische Physik and Riemann Center for Geometry and Physics \\
Leibniz Universit\"at Hannover,
Appelstrasse 2, 30167 Hannover, Germany}

\vspace{2mm}

${}^c$ {\it Yerevan Physics Institute, 2 Alikhanian Brothers Street, 0036 Yerevan, Armenia}
\vspace{5mm}

{\tt nkozyrev@theor.jinr.ru, krivonos@theor.jinr.ru, lechtenf@itp.uni-hannover.de, arnerses@ysu.am, sutulin@theor.jinr.ru}

\end{center}

\vspace{2cm}

\begin{abstract}\noindent
We present ${\cal N}{=}\,4$ supersymmetric mechanics on $n$-dimensional
Riemannian manifolds constructed within the Hamiltonian approach.
The structure functions entering the supercharges and the Hamiltonian
obey modified covariant constancy equations as well as
modified Witten--Dijkgraaf--Verlinde--Verlinde equations specified
by the presence of the manifold's curvature tensor.
Solutions of original Witten--Dijkgraaf--Verlinde--Verlinde equations
and related prepotentials defining ${\cal N}{=}\,4$ superconformal mechanics in flat space
can be lifted to $so(n)$-invariant Riemannian manifolds.
For the Hamiltonian this lift generates an additional potential term which,
on spheres and (two-sheeted) hyperboloids, becomes a Higgs-oscillator potential.
In particular, the sum of $n$ copies  of one-dimensional conformal mechanics results in
a specific superintegrable deformation of the Higgs oscillator.
\end{abstract}

\newpage
\pagenumbering{arabic}
\setcounter{page}{1}

\section{Introduction}
The Witten--Dijkgraaf--Verlinde--Verlinde (WDVV) equations were introduced a few decades ago in the context of two-dimensional topological field theories \cite{WDVV}.
In their initial version they read
\be\label{flatWDVV}
 F^{(0)}_{ijm}\delta^{nm} F^{(0)}_{kln}=F^{(0)}_{ilm}\delta^{mn} F^{(0)}_{kjn}\qquad{\rm with }\quad F^{(0)}_{ijk}=\partial_i \partial_j \partial_k F^{(0)}(x).
\ee
Clearly, these equations are not covariant with respect to general coordinate transformations, so their covariantization requires introducing additional
geometric structures. Finding solutions to the WDVV equations is a nontrivial task, which was considered in numerous papers in various contexts
 (see, e.g. \cite{Dubrovin, WDVVsol1, WDVVsol2, WDVVsol3, WDVVsol4, WDVVsol5,  WDVVsol6} and refs.\ to them).
The  WDVV equations appear  also when constructing  ${\cal N}{=}\,4$ supersymmetric/superconformal  extensions of  mechanics on $n$-dimensional Euclidean space
(see \cite{Wyllard, BGL, GLP, GLP1, KLP, KL1} and refs.\ therein). Indeed, as it was firstly demonstrated in \cite{Wyllard}, if we insist that the
${\cal N}{=}\,4$ supercharges $Q^a$ and $\bQ_b$ with $a,b =1,2$, in the form
\be\label{Qf}
Q^a = p_i \,\psi^{a i } + \im W^{(0)}_i\, \psi^{a i}+\im F^{(0)}_{ijk}\, \psi^{b i}\, \psi^j_{b}\, \bpsi^{a k}
\qquad\textrm{and}\qquad
\bQ_a = p_i\, \bpsi^i_a  - \im W^{(0)}_i\, \bpsi^i_a +\im F^{(0)}_{ijk}\, \bpsi^i_b\, \bpsi^{b j}\, \psi^k_a,
\ee
obey the ${\cal N}{=}\,4$ super Poincar\'{e} algebra
\be\label{N4P}
\left\{ Q^a, Q^b\right\}=0, \qquad
\left\{\bQ_a, \bQ_b \right\}=0, \qquad
\left\{ Q^a, \bQ_b\right\} =  \sfrac{\im}{2} \, \delta^a_b \,H ,
\ee
then the totally symmetric structure functions $F^{(0)}_{ijk}(x)$ entering the supercharges \p{Qf} have to satisfy \p{flatWDVV}, while the prepotential $W^{(0)}$ is found by solving
\be\label{fpot}
\partial_i \partial_j W^{(0)}  + F^{(0)}_{ijk}\partial_k W^{(0)}=0
\qquad\textrm{and}\qquad W^{(0)}_i = \partial_i W^{(0)}.
\ee
We should add that, when evaluating the brackets between the supercharges in \p{N4P}, the basic variables were taken to obey the standard Dirac brackets
\be\label{DB1}
\big\{ x^i, {p}{}_j   \big\}= \delta^{i}_j \qquad\textrm{and}\qquad
\big\{ {\psi}^{a i}, {\bpsi}_b^j    \big\} = \sfrac{\im}{2}\delta^a_b \delta^{ij}.
\ee

Exploiting this relation, one can postulate a general coordinate invariant version of the WDVV equations
simply by constructing ${\cal N}{=}\,4$ supersymmetric mechanics on arbitrary Riemannian spaces.
This was achieved in our recent paper~\cite{cWDVV}.
To construct such mechanics, one has to generalize
the ansatz for the supercharges \p{Qf} and the Poisson brackets \p{DB1} to be covariant under general coordinate transformations, and then to check the
conditions on the structure functions implied by the ${\cal N}{=}\,4$ super-Poincar\'{e} algebra relations \p{N4P} these supercharges should
obey. While \cite{cWDVV} accomplished this for the special case of $W_i=0$ -- no potential term --
its result for the structure functions $F_{ijk}$ remains true in general:
\be \label{mWDVV}
\nabla_i F_{jkm}= \nabla_j F_{ikm} \qquad\textrm{and}\qquad
F_{ikp}g^{pq}F_{jmq}-F_{jkp}g^{pq}F_{imq}+R_{ijkm}=0.
\ee
These are the curved WDVV equations on spaces with a metric $g_{ij}$.
It was also shown there that solutions of the flat WDVV equations \p{flatWDVV}
can be extended to solutions of the curved WDVV equations~\p{mWDVV} on  isotropic
spaces.

The main goal of the present paper is to construct ${\cal N}{=}\,4$ supersymmetric mechanics with a nonzero potential on arbitrary
Riemannian spaces.
As a first step, in Section~2 we introduce generalized Poisson brackets which are covariant with
respect to general coordinate transformations. Then we write down the most general ansatz for the supercharges (linear and cubic in the fermionic variables, for the case with and without additional
spin variables) and analyze the conditions on the structure functions. We obtain equations on the prepotentials
which generalize \p{fpot}. In Section~3 we specialize to isotropic spaces and extend the general
solution found in~\cite{cWDVV} to include the prepotential~$W_i$.
In Section~4 we
present some explicit solutions for the most interesting spaces -- spheres and pseudospheres.
Finally, we conclude with a few comments and remarks.

\newpage

\setcounter{equation}{0}
\section{Symplectic structure, supercharges and Hamiltonian}
We are going to construct the supercharges $Q^a, \bQ{}_b,\; (a,b=1,2)$ 
forming an ${\cal N}{=}\,4, d=1$ Poincar\'{e} superalgebra\footnote{
We use the following convention for raising and lowering $su(2)$ indices: 
$\cA_a =\epsilon_{ab}\cA^b, \cA^a= \epsilon^{ab} \cA_b, \; \epsilon_{12}=\epsilon^{21}=1$.}
\be\label{N4Poincare}
\left\{ Q^a, Q^b\right\}=0, \qquad
\left\{\bQ_a, \bQ_b \right\}=0, \qquad
\left\{ Q^a, \bQ_b\right\} =  \sfrac{\im}{2} \, \delta^a_b \,H
\ee
for $n$-dimensional systems in which each bosonic degree of freedom $x^i\; (i=1,\ldots, n)$ is accompanied by four fermionic ones $\psi^{a i}, \bpsi^j_{b}=\left( \psi^{b j}\right)^\dagger$. The extended phase space, parameterized by the bosonic coordinates $x^i$ and momenta $p_i$ and the fermionic coordinates $\psi^{a i}, \bpsi^j_{b}$ , can be equipped by the symplectic structure
\be\label{ss}
\Omega=dp_i\wedge dx^i+
{\im}d\left(\psi^{ia}g_{ij}{ D}\bpsi^{j}_{a}- \bpsi^{ia}g_{ij}{ D} \psi^{j}_{a} \right)=  dp_i \wedge dx^i +
 \im R_{ijkl}\psi^{ia}\bpsi^{j}_{a} dx^k\wedge
dx^l +2\im g_{ij}D\psi^{ia} \wedge D\bpsi^{j}_{a},
\ee
where $D\psi^{ia}\equiv d\psi^{ia}+ \Gamma^i_{jk}\psi^{ja} dx^k$,
and    $\Gamma^i_{jk}$ and $R^i_{jkl}$  are the components of the Levi-Civita connection and curvature  of the metric  $g_{ij}(x)$ defined in a standard way as
\be\label{def1}
\Gamma^{k}_{ij} = \sfrac{1}{2} \; g^{km} \left( \partial_i g_{jm}+ \partial_j g_{im} - \partial_m g_{ij} \right), \qquad\textrm{and}\qquad
R^i{}_{jkl} = \partial_k \Gamma^i_{j l} - \partial_l \Gamma^i_{j k}+ \Gamma^m_{j l}\; \Gamma^i_{m k}-\Gamma^m_{j k}\; \Gamma^i_{m l}.
\ee
This symplectic structure  is manifestly  invariant with respect to the transformations
\be
{\tilde x}^i={\tilde x}^i(x),\qquad
{\tilde p}_i=\frac{\partial x^j}{\partial \tilde x^i}p_j, \qquad
{\tilde\psi^{ia}}=\frac{\partial{\tilde x}^i(x)}{\partial x^j}\psi^{ja}.
\label{LT}
\ee
The Poisson brackets between the basic variables can be immediately extracted from the symplectic structure \p{ss} :
\bea\label{PB}
&& \big\{ x^i, {p}{}_j   \big\}= \delta^{i}_j, \;\; \big\{ {\psi}^{ai}, {\bpsi}_b^j    \big\} = \frac{\im}{2}\delta^a_b \, g^{ij}, \;\;
\big\{ {p}_i, { p}_j \big\} = -2\im R_{ijkm}{\psi}^{a k}{\bpsi}_a^m, \nn \\
&& \big\{ {p}_i, {\psi}^{aj} \big\} = \Gamma^j_{ik}{\psi}^{ak}, \; \; \big\{ { p}_i, {\bpsi}_a^{j} \big\} = \Gamma^j_{ik}{\bpsi}_a^{k}.
\eea

To construct the supercharges $Q^a, \bQ{}_b$ we have two possibilities.
\begin{itemize}
\item One may construct the standard supercharges in terms of the variables $x^i, p_j,  \psi^{a i}, \bpsi^j_{b}$ only, mainly following to the line of the paper \cite{GLP}.
\item One may extend the set of the basic variables by the additional bosonic spin variables $\left\{ u^a, \bu_a|a=1,2\right\}$ parameterizing an internal two-sphere and obeying the brackets
\be\label{PBu}
\left\{ u^a, \bu_b \right\} = - \im \, \delta^a_b .
\ee
These new variables will appear in the supercharges only through the $su(2)$ currents \cite{FIL,KL1}
\be\label{su2}
J^{ab} = \sfrac{\im}{2} \left( u^a \bu^b +u^b \bu^a\right) \quad \Rightarrow \quad
\left\{ J^{ab} ,J^{cd} \right\} = -\epsilon^{ac} J^{bd} - \epsilon^{bd} J^{ac} .
\ee
\end{itemize}
Let us consider these possibilities separately.

\subsection{Standard supercharges}
The most general ansatz  for the standard supercharges reads:
\bea\label{Q}
&& Q_W^a = p_i \,\psi^{a i } + \im W_i\, \psi^{a i} +\im F_{ijk}\, \psi^{b i}\, \psi^j_{b}\, \bpsi^{a k}+ \im G_{ijk}\,\psi^{a i}\, \psi^{b j} \,\bpsi_b^{k}, \nn\\
&& \left( \bQ_W \right)_a = p_i\, \bpsi^i_a  - \im W_i\, \bpsi^i_a +\im F_{ijk}\, \bpsi^i_b\, \bpsi^{b j}\, \psi^k_a + \im G_{ijk}\,\bpsi^i_a \, \bpsi^j_{b}\, \psi^{b k} \;.
\eea
Here, $W_i, F_{ijk}$ and $G_{ijk}$ are arbitrary, for the time being, real functions depending on $n$ coordinates $x^i$.
In addition, we assume that the functions $F_{ijk}$ and $G_{ijk}$ are symmetric and anti-symmetric over the first two indices, respectively:
\be
F_{ijk} =F_{jik}, \qquad G_{ijk} = - G_{jik}.
\ee
The conditions that these supercharges span an ${\cal N}{=}\,4$ super Poincar\'{e} algebra \p{N4Poincare} result in the following equations on the functions involved:
\bea
&& G_{ijk} =0, \qquad F_{ijk}-F_{ikj} =0\qquad\Rightarrow\qquad
\textrm{ $F_{ijk}$ is totally symmetric,} \label{cond1a} \\
&& \nabla_i F_{jkm}- \nabla_j F_{ikm}=0, \label{cond1b} \\
&&  F_{ikp}g^{pq}F_{jmq}-F_{jkp}g^{pq}F_{imq}+R_{ijkm}=0, \label{cond1c}
\eea
and
\bea
&& \nabla_i W_j - \nabla_j W_i =0 \qquad\Rightarrow\qquad W_i = \partial_i W , \label{cond1d} \\
&& \nabla_i \partial_j W + F_{ijk} g^{km} \partial_m W =0, \label{cond1e}
\eea
where, as usual,
\bea
&& \nabla_i W_j = \partial_i W_j - \Gamma^k_{ij} W_k, \nn \\
&& \nabla_i F_{jkl} = \partial_i F_{jkl} -\Gamma^m_{ij} F_{klm}-\Gamma^m_{ik} F_{jlm}-\Gamma^m_{il} F_{jkm}.
\eea
Once the equations \p{cond1a}-\p{cond1e} are satisfied, the Hamiltonian $H$ acquires form
\be\label{Ham1}
H_W= g^{ij} { p}{}_i { p}{}_j + g^{ij} \partial_i W \partial_j W +4  \nabla_i \partial_j W {\psi}^{ci} {\bpsi}_{c}{}^j -4\big[  \nabla_m F_{ijk} +  R_{imjk} \big] {\psi}^{ci}  {\bpsi}_{c}{}^m\,{\psi}^{dj} {\bpsi}_{d}{}^k.
\ee

\subsection{Supercharges with spin variables}
Following \cite{FIL,KL1}, the spin variables $\left\{ u^a, \bu_a\right\}$ may be utilized to slightly modify the prepotential term in the supercharges to be
\bea\label{Q2}
&& Q_U^a =  p_i \, \psi^{i a} + U_i\, J^{ab}\,\psi^i_{b} -\im F_{ijk}\, \psi^{b i}\, \psi^j_{b}\, \bpsi^{a k}+ \im G_{ijk}\,\psi^{a i}\, \psi^{b j} \,\bpsi_b^{k}, \nn \\
&& \left(\bQ_U\right)_a = p_i \, \bpsi_a^{i} - U_i\, J_{ab}\,\bpsi^{b i} - \im F_{ijk}\, \bpsi^i_b\, \bpsi^{b j}\, \psi^k_a + \im G_{ijk}\,\bpsi^i_a \, \bpsi^j_{b}\, \psi^{b k} \;.
\eea
These supercharges form an ${\cal N}{=}\,4$ super Poincar\'{e} algebra if the functions $F_{ijk}$ and $G_{ijk}$ obey
the same constraints \p{cond1a} - \p{cond1c}, while the constraints \p{cond1d}, \p{cond1e} which include the prepotential are changed to
\bea
&& \nabla_i U_j - \nabla_j U_i =0 \qquad\Rightarrow\qquad U_i = \partial_i U , \label{cond2d} \\
&& \nabla_i \partial_j U -\partial_i U \partial_j U + F_{ijk} g^{km} \partial_m U =0. \label{cond2e}
\eea
When the constraints \p{cond1a} - \p{cond1c} and \p{cond2d}, \p{cond2e} are satisfied, the Hamiltonian reads
\be\label{Ham2}
H_U= g^{ij} { p}{}_i { p}{}_j + \sfrac{1}{2} J^{ab} J_{ab} g^{ij} \partial_i U \partial_j U - 4 \im  \nabla_i \partial_j U J_{ab} {\psi}^{ai} {\bpsi}^{b j} +4\big[ \nabla_m F_{ijk} -  R_{imjk} \big] {\psi}^{ci}  {\bpsi}_{c}{}^m\,{\psi}^{dj} {\bpsi}_{d}{}^k.
\ee
As we can see now, the $su(2)$ Casimir element $J_{ab}J^{ab}$ plays the role of the coupling constant.

Finally, note that  the solutions of the equations \p{cond1d}, \p{cond1e} and \p{cond2d}, \p{cond2e} are related as follows,
\be\label{harm}
W\;=\;\textrm{e}^{- U}.
\ee
Thus, any solution of the system \p{cond1d}, \p{cond1e} generates a solution of the system  \p{cond2d}, \p{cond2e}, and vice versa.
It should be stressed, however, that the exact form of the bosonic potentials and terms quadratic in fermionic variables  are very different in
the Hamiltonians $H_W$ \p{Ham1} and $H_U$ \p{Ham2}.

Equation \p{cond1b} qualifies $F_{ijk}$ as a so-called third-rank Codazzi tensor \cite{CodazziT}, while \p{cond1c} is the ‘curved WDVV
equation’ proposed in \cite{cWDVV}, for which the equations  \p{cond1e}, \p{cond2e} are the curved analogs of the flat equations on the
prepotentials discussed in \cite{GLP} and \cite{KL1}.

Summarizing, one may conclude that to construct ${\cal N}{=}\,4$ supersymmetric $n$-dimensional  mechanics with a given bosonic metric $g_{ij}$ one has
\begin{itemize}
\item To solve the curved WDVV equations \p{cond1b}, \p{cond1c} for the fully symmetric function $F_{ijk}$,
\item To find  admissible prepotentials as  solutions of the equations \p{cond1d}, \p{cond1e} and/or \p{cond2d}, \p{cond2e}.
\end{itemize}
In what follows, we will use this procedure to construct ${\cal N}{=}\,4$ supersymmetric $n$-dimensional mechanics.

\newpage

\setcounter{equation}{0}
\section{Isotropic spaces}
\subsection{Solution of curved WDVV equation}
In \cite{cWDVV} a large class of solutions to the equations \p{cond1b}, \p{cond1c}  has been constructed on isotropic spaces. Such spaces admit an $so(n)$-invariant metric with components
\be\label{CFlat1}
g_{ij} = \frac{1}{f(r)^2} \delta_{ij}, \quad \Gamma^k_{ij} =-\frac{f'}{ r f} \left( x^i \delta_j^k+x^j\delta_i^k-x^ k \delta_{ij}\right),
\quad r^2 = \delta_{ij} x^i x^j.
\ee
The key point of the analysis performed in \cite{cWDVV} is the following ansatz on the structure function $F_{ijk}$:
\be\label{CFanz1}
F_{ijk} = a(r)  x^{i} x^{j} x^{k} + b(r) \left( \delta_{ij}  x^{k} +
 \delta_{jk}  x^{i} +\delta_{ik} x^{j}\right)+  f(r)^{-2} F^{(0)}_{ijk},
\ee
where $F^{(0)}_{ijk}$ is an arbitrary solution of the flat WDVV equation, i.e.
\be\label{WDVV}
F^{(0)}_{ikp}\delta^{pq}F^{(0)}_{jmq}-F^{(0)}_{jkp}\delta^{pq}F^{(0)}_{imq}=0 \qquad \mbox{with} \qquad F^{(0)}_{ijk}=\partial_i \partial_j \partial_k F^{(0)}.
\ee
One may check that the linear equation \p{cond1b} is satisfied if
\be\label{mWDVV1}
r ( r f' -f ) a + 4 f' b + f b'=0 \qquad \mbox{and} \qquad
x^i F^{(0)}_{ijk} = c \delta_{jk}, \quad c= \textrm{const},
\ee
where we fixed the scale of $F^{(0)}$. Here, prime means differentiation with respect to  $r$. Thus, the symmetric tensor $F_{ijk}$ \p{CFanz1} with the restrictions
\p{WDVV}, \p{mWDVV1} provides a third rank Codazzi tensor on isotropic spaces  \p{CFlat1}.

The curved WDVV equation \p{cond1c} further imposes the quadratic conditions
\be\label{finequ}
f^2 b \left( r^2 a +b \right) + c a = -\frac{1}{r f^3} \left( \frac{f'}{r}\right)' \qquad \mbox{and} \qquad
r^2 f^2 b^2 + 2 c b = \frac{r^2 f'}{f^4} \left( \frac{f}{r^2}\right)' .
\ee
We note that these equations already imply the condition \p{mWDVV1}.

The equations \p{finequ} may be easily solved as
\be \label{solmWDDV}
\begin{aligned}
a &\= \frac{2 c f \sqrt{c^2 f^2 - 2r f f' + r^2(f')^2} \pm \bigl( 2c^2f^2 - 3r f f' + r^2(f')^2 + r^2 f f''\bigr)}
{r^4 f^3 \sqrt{c^2 f^2 - 2r f f' + r^2(f')^2}} \ ,\\
b &\= -\frac{c f \pm \sqrt{c^2 f^2 - 2r f f' + r^2(f')^2}}{r^2 f^3} \ .
\end{aligned}
\ee
For $c{=}1$, it simplifies to
\be\label{solWDVV2}
a \= \frac{2 f \bigl( f- r f'\bigr) \pm \bigl( 2 f^2 - 3 r f f'+ r^2 (f')^2+ r^2 f f''\bigr)}{r^4 f^3 \bigl( f - r f'\bigr)}
\quad\und\quad
b \= -\frac{f \pm \bigl( f - r f'\bigr)} {r^2 f^3} \ .
\ee

It should be noted that the solution for $a$ in \p{solmWDDV} becomes 0/0 indeterminate if
\be\label{expoint1}
f_0(r) = \mu r^{1 \pm \sqrt{1-c^2}}, \quad \mu = \textrm{const}.
\ee
With such a metric the solution of the equations \p{mWDVV1}, \p{finequ} reads
\be\label{expoint1a}
a = \frac{2 c}{ r^4 f_0^2}, \quad b=- \frac{c}{ r^2 f_0^2}.
\ee
Another exceptional case corresponds to  $c=\pm 1$ and the metric function
\be\label{expoint2}
f_1(r) = \mu r.
\ee
For this case, the function $a(r)$ is not restricted while $b(r)$ has the form
\be\label{expoint2a}
b=- \frac{c}{\mu^2 r^4}.
\ee

\subsection{Searching for the prepotentials}
Having at hands the solution for the curved WDVV equations \p{cond1b}, \p{cond1c}, one may try to solve the equations for the prepotentials \p{cond1d}, \p{cond1e} or
\p{cond2d}, \p{cond2e}, respectively. It should be noted that even in the flat case, where a variety of the solutions to the WDVV equations is known
\cite{WDVVsol1, WDVVsol2, WDVVsol3, WDVVsol4, WDVVsol5}, such a task is far from being completely solved. Nevertheless, many particular solutions are known for the
standard supercharges \cite{GLP, GLP1, KLP} as well as for the case with spin variables \cite{KL1}. Leaving the full analysis of the admissible prepotentials
for the future, let us demonstrate that each prepotential found for the flat WDVV equations can be embedded into isotropic spaces with a metric \p{CFlat1}.

\subsubsection{Lifting a ``flat'' prepotential $W^{(0)}$}
In this case we have to solve the equation \p{cond1d} which for the metric \p{CFlat1} and for the $F_{ijk}$ given in \p{CFanz1} acquires the form
\be\label{pot1}
 \partial_i \partial_j W +\left( \frac{f'}{r f} +b f^2 \right)\left( x^{i} \partial_j W+  x^{j} \partial_i W\right)+ \delta_{ij}\left( b f^2  - \frac{f'}{r f}\right)  x^m \partial_m W
 + a f^2   x^{i}  x^{j} x^m  \partial_m W+ F^{(0)}_{ijm}\delta^{mn}\partial_n W=0.
\ee
Let us also suppose that we know the solution $W^{(0)}$ of the flat equation
\be\label{flatpot}
\partial_i \partial_j W^{(0)}  + F^{(0)}_{ijm}\delta^{mn}\partial_n W^{(0)}=0.
\ee
All such solutions found in \cite{GLP, GLP1, KLP} obey the additional condition
\be\label{addcond}
x^i \partial_i W^{(0)} = \alpha = \textrm{const} \qquad \Rightarrow \qquad x^m F^{(0)}_{m j k} = \delta_{jk},
\ee
i.e. the parameter $c$ in \p{solmWDDV} fixed to be equal to one.

If we now choose the following ansatz for the prepotential $W$,
\be
W = \tW(r) + W^{(0)},
\ee
then the ``flat'' prepotential $W^{(0)}$ will appear in the equation \p{pot1} only through the constant $\alpha$, except for the second term
$$
\sim \left( \frac{f'}{r f} +b f^2 \right)\left( x^{i} \partial_j W^{(0)}+  x^{j} \partial_i W^{(0)}\right).
$$
To kill this term we have to choose $b = - \frac{f'}{r f^3}$. This choice corresponds to the following solution in \p{solWDVV2}
\be\label{sol11}
a = \frac{ f f' - r (f')^2 -r f f''}{r^3 f^3 \left( f - r f'\right)}, \quad b = - \frac{f'}{r f^3}.
\ee
Finally, it is a matter of straightforward calculations to check that the prepotential
\be\label{solpot1}
W = \tW(r) + W^{(0)} \qquad \mbox{with}\qquad \tW' = \alpha \frac{f'}{f-r f'}
\ee
solves the equation \p{pot1}. Correspondingly, the bosonic potential in the Hamiltonian \p{Ham1} reads
\be\label{ham1pot}
 g^{ij} \partial_i W \partial_j W = f^2\left[ \delta^{ij} \partial_i W^{(0)} \partial_j W^{(0)} +
 \alpha^2 \frac{f'\left( 2 f- r f'\right)}{r \left( f- r f'\right)^2}\right].
\ee

\subsubsection{Lifting a ``flat'' prepotential with spin variables}
For the supercharges with spin variables the solutions for the equation \p{cond2e} may be found by using the relation \p{harm}. However, the additional constraint
for the flat prepotential \p{addcond} we used above is a quite unconventional. Indeed, in  \cite{KL1} the additional condition on the solution of the flat equation
\be\label{flat11}
\partial_i \partial_j U^{(0)} - \partial_i U^{(0)} \, \partial_j U^{(0)} + F^{(0)}_{ijk}\delta^{km}\partial_m U^{(0)}=0
\ee
reads
\be\label{condtU}
x^k \partial_k U^{(0)}= \alpha -1 \qquad \Rightarrow \qquad x^k F^{(0)}_{ijk}= \alpha\;\delta_{ij}.
\ee
Therefore, we need to reconsider the solution of  \p{cond2e} using an ansatz
\be\label{Uans}
U = \tU(r)  + U^{(0)}.
\ee
We will use the same ansatz for the $F_{ijk}$ \p{CFanz1} with the same conditions on the functions $a(r) , b(r)$ \p{finequ} and the constraint \p{condtU}.

Now, it is rather easy  to check that the prepotential $U$ \p{Uans} obeys the equations \p{cond2e} if
\be\label{Ueq3sol}
\tU^\prime = \frac{f^\prime}{f}+ r f^2 b.
\ee
Therefore, the resulting potential in the Hamiltonian \p{Ham2} reads
\be\label{Upot}
V =\frac{1}{2} J^{ab}J_{ab} f^2 \left[ \delta^{ij}\partial_i U^{(0)} \partial_j U^{(0)}  +\frac{2(\alpha-1)}{r} \left(\frac{f^\prime}{f}+ r f^2 b\right) +
\left(\frac{f^\prime}{f}+ r f^2 b\right)^2
 \right],
\ee
with $b$ given in \p{solmWDDV}.

\subsubsection{Lifting a vanishing ``flat'' prepotential}
This case corresponds to the absence of the ``flat'' prepotential, i.e.\ to the case with $W^{(0)}=0$. To simplify the analysis we suppose that the prepotential $W$ depends
on $r$ only, while the ``flat''
WDVV solution $F^{(0)}_{ijk}$ still obeys the constraint \p{addcond}, i.e.
\be
x^i F^{(0)}_{ijk} = \delta_{jk} \qquad \Rightarrow \qquad c=1.
\ee
With these assumptions the equation \p{cond1d} reads
\be\label{pot2}
\delta^{ij} \frac{2 f + b r^2 f^3 - r f'}{r f} W' + x^i x^j
\frac{ r f W'' + \left( -f + \left( 2 b + a r^2\right) r^2 f^3 + 2 r f'\right)W'}{r^3 f} =0.
\ee
To kill the first term in \p{pot2} we have to choose $b=\frac{- 2 f +r f'}{r^2 f^3}$. This choice corresponds to the following solution in \p{solWDVV2},
\be\label{sol12}
a = \frac{-r f' +f \left( 4+ \frac{r^2 f''}{f-r f'}\right)}{r^4 f^3}, \quad b =  \frac{- 2 f +r f'}{r^2 f^3}.
\ee
With such functions $a, b$  the  equation which defines the prepotential
acquires the form
\be\label{pot22}
r W'' + \left( -1 + \frac{3 r f'}{f}+\frac{ r^2 f''}{f- r f'}\right) W'=0.
\ee
This equation has the general solution
\be\label{solPot2}
W = c_1\left( \frac{r^2}{f^2} + c_2\right), \qquad c_1,c_2 = \textrm{const}.
\ee
The bosonic potential in the Hamiltonian \p{Ham1} now reads
\be\label{ham2pot}
 g^{ij} \partial_i W \partial_j W = 4\;c_1^2 \;\frac{r^2}{f^4}\; \left( f - r f'\right)^2.
\ee

It is interesting that, after passing to the prepotential $U=-\log(W)$ \p{harm}, the potential term in the Hamiltonian with spin variables \p{Ham2}
acquires the form
\be
\sfrac{1}{2} J^{ab} J_{ab} g^{ij} \partial_i U \partial_j U = J^{ab} J_{ab}\; \frac{ 2 r^2 \left( f - r f'\right)^2}{\left( r^2 + c_2 f^2\right)^2}.
\ee
Thus, we see that the constant $c_2$ entering the solution \p{solPot2} and having no impact on the Hamiltonian $H_W$ \p{ham2pot} becomes quite
important for the Hamiltonian $H_U$ \p{Ham2}.

\setcounter{equation}{0}
\section{Examples of prepotentials on the (pseudo-)sphere}
\subsection{Supersymmetric black holes}
In this section we present some interesting prepotentials on (pseudo-)spheres which admit ${\cal N}{=}\,4$ supersymmetry.
As we can see from the previous section, any ``flat'' system obeying the constraints \p{addcond}, i.e.
\be
x^i \partial_i W^{(0)} = \alpha = \textrm{const} \qquad \Rightarrow \qquad x^m  F^{(0)}_{m j k} = \delta_{jk},
\ee
has its image on isotropic spaces with the potential \p{ham1pot}
\be
\cV =g^{ij} \partial_i W \partial_j W =
 f^2\left[ \delta^{ij} \partial_i W^{(0)} \partial_j W^{(0)} + \alpha^2 \frac{f'\left( 2 f- r f'\right)}{r \left( f- r f'\right)^2}\right].
\ee
In the case of a (pseudo-)sphere with the metric
\be\label{sphere}
f = 1+\epsilon r^2, \qquad \epsilon = \pm 1,
\ee
the potential $\cV$ is simplified to be
\be\label{potSPH1}
\cV =\left( 1+\epsilon r^2\right)^2  \delta^{ij} \partial_i W^{(0)} \partial_j W^{(0)} + 4 \alpha^2 \epsilon \left( \frac{1+\epsilon r^2}{1-\epsilon r^2}\right)^2.
\ee
Thus we see that, besides getting multiplied with the standard factor $\left( 1+\epsilon r^2\right)^2$, the ``flat'' potential is shifted by the potential of a Higgs
oscillator \cite{higgs},
\be\label{HP}
V_{\textrm{Higgs}} \equiv \left( \frac{1+\epsilon r^2}{1-\epsilon r^2}\right)^2.
\ee
This means that even a mutually non-interacting system, being placed on the (pseudo-)sphere, becomes interacting via the Higgs potential.
A prominent example comes from the sum of several ${\cal N}{=}\,4$ supersymmetric mechanics on flat space with conformal prepotentials
\be\label{ex1a}
W^{(0)} = \sum_i^n a_i \log(x^i), \qquad F^{(0)}= \frac{1}{2}\sum_i^n (x^i)^2 \log(x^i).
\ee
With such almost ``free'' prepotentials  we  obtain the following potential for the system on the (pseudo-)sphere,
\be\label{ex1b}
\alpha = \sum_i^n a_i \qquad \Rightarrow \qquad \cV_1=\left(1+\epsilon r^2\right)^2 \left[ \sum_i^n \frac{a_i^2}{(x^i)^2}+
\frac{ 4\epsilon \left(\sum_i^n a_i\right)^2}{\left( 1 - \epsilon r^2 \right)^2}\right].
\ee
A system with such a potential was obtained as a reduced angular (compact) part of conformal mechanics describing the motion of a
particle in a near-horizon Myers--Perry black-hole background with coinciding rotational parameters in \cite{Armen1}, and
its superintegrability was proven there as well. 
%Note that ${\cal N}{=}\,4$ supersymmetry imposes an additional constraint on the coupling constants, namely $a_{n+1} = \sum_i^n a_i$.

\subsection{(Pseudo-)sphere image of a free system}
This case is analogous to the one previously considered but, unfortunately, we cannot just substitute $W^{(0)}=0$ into \p{potSPH1} because this is in
contradiction with the constraint $x^i \partial_i W^{(0)}=\alpha$. Thus, we have to use the general consideration in Subsection 3.2.3.

For the (pseudo-)sphere with  metric $f=1+\epsilon r^2$ the potential in \p{ham2pot} reads
\be\label{ex3a}
{\cal V}_W = 4 c_1^2 \frac{r^2 \left(1-\epsilon r^2 \right)^2}{\left(1+\epsilon r^2\right)^4}.
\ee
At the same time, the potential for the supercharges with spin variables can be easily obtained by passing to the prepotential $U$ \p{harm},
\be\label{ex3a1}
U = -\log(W) = -\log\left[ c_1 \left( \frac{r^2}{(1+\epsilon r^2)^2}+c_2\right)\right],
\ee
and by evaluating the potential for the case with spin variables in the Hamiltonian \p{Ham2} we get
\be\label{ex3a2}
\cV_U= \sfrac{1}{2}\; J^{ab} J_{ab}\; g^{ij} \partial_i U \partial_j U= 2 \;J^{ab} J_{ab}\;
\frac{r^2 \left(1-\epsilon r^2\right)^2}{\left(r^2 +c_2 \left( 1+\epsilon r^2\right)^2\right)^2}.
\ee
Choosing now
\be
c_2 = -\frac{1}{4\epsilon}
\ee
we obtain
\be\label{ex3a3}
\cV_U= 32\;\epsilon^2\; J^{ab} J_{ab}\;  \frac{r^2}{\left(1-\epsilon r^2\right)^2} =8 \epsilon\; J^{ab} J_{ab}\;
\left[ \frac{\left(1+\epsilon r^2\right)^2}{\left(1-\epsilon r^2\right)^2}-1\right] \;\equiv\; 8 \epsilon\; J^{ab} J_{ab}\; \left[ V_{\textrm{Higgs}} -1\right].
\ee
Thus, the Higgs oscillator in the Hamiltonian with spin variables on the (pseudo-)sphere is the image of a completely free system.

\subsection{(Pseudo-)sphere image of the isotropic harmonic oscillator}
Due to the exceptional role the standard harmonic oscillator  plays among integrable systems, it is interesting to find its image in
the case of ${\cal N}{=}\,4$ supersymmetric mechanics on the (pseudo-)sphere. Unfortunately, in this case our consideration of the previous sections does not
help too much because, if we choose
\be\label{ho1}
W^{(0)} = m r^2 \qquad\textrm{with}\quad m= \textrm{const},
\ee
then the condition $x^i \partial_i W^{(0)}=\alpha$ we used previously will not be valid anymore.
However, plugging the expression \p{ho1} into \p{flatpot} we will get
\be\label{ho2}
\partial_i \partial_j W^{(0)}  + F^{(0)}_{ijm}\delta^{mn}\partial_n W^{(0)}=0 \quad \Rightarrow \quad x^m  F^{(0)}_{ijm}= - \delta_{ij}.
\ee
Thus, in the general solution \p{CFanz1}, \p{solmWDDV} for the curved WDVV equation  we have to substitute $c=-1$. Thus, everything is simplified, and we have two solutions for the
(pseudo-)sphere with $f= 1+\epsilon r^2$:
\bea
&& a_1 = \frac{ 4 \epsilon^2}{\left(1- \epsilon r^2\right)\left(1+ \epsilon r^2\right)^3}, \quad b_1 = \frac{2\epsilon}{\left(1+ \epsilon r^2 \right)^3}, \label{sol1}\\
&& a_2 = -\frac{ 4 }{r^4\left(1- \epsilon r^2\right)\left(1+ \epsilon r^2\right)^3}, \quad b_2 = \frac{2}{r^2 \left(1+ \epsilon r^2 \right)^3} \label{sol2}.
\eea
Now, the basic equation which defines the admissible prepotentials \p{pot1} for the prepotentials $\tW=\tW(r)$ acquires the form
\bea
0 &=& x^i x^j \left[ \frac{\tW''}{r^2}- \frac{\tW'}{r^3} +
\frac{4 \epsilon + \left(2 b+ a r^2\right) \left( 1+ \epsilon r^2\right)^3}{r \left( 1+ \epsilon r^2\right)}  \left( \tW'+ 2 m r\right)
\right] \nn \\
&& +\ \delta_{ij} r \left( 1+ \epsilon r^2\right)^2 \Bigl( b  - \frac{2 \epsilon}{\left( 1+ \epsilon r^2\right)^3}\Bigr)\left( \tW'+ 2 m r\right).
\eea
Now we see that, to kill the $\delta_{ij}$ term, we have to choose the first solution \p{sol1}. Therefore, the equation for the prepotential $\tW$ reads
\be\label{fineqW}
r \left( 1- \epsilon r^2\right)\left( 1+ \epsilon r^2\right)\tW'' -\left( 1 - 8\epsilon r^2+3 \epsilon^2 r^4\right) \tW'+ 8 m \epsilon r^3 \left( 2- \epsilon r^2\right)=0,
\ee
with the solution
\be\label{finpot1}
\tW = - \frac{ 2 m \bigl( 1 + 15 \epsilon r^2+3 \epsilon^2 r^4+\epsilon^3 r^6\bigr) +\epsilon r^2 M_1}{2 \epsilon \bigl( 1+ \epsilon r^2\bigr)^2}+M_2,
\ee
where $M_1$ and $M_2$ are integration constants. Thus, the (pseudo-)sphere image of the oscillator potential $W^{(0)}$ \p{ho1} reads
\be\label{ho3}
\cV_W = f^2 \delta^{ij} \partial_i \left( \tW+W^{(0)}\right) \partial_j \left( \tW+W^{(0)}\right) =
\left( 24 m + M_1\right)^2 r^2 \frac{ \left( 1 - \epsilon r^2\right)^2}{ \left( 1 + \epsilon r^2\right)^4}.
\ee
One may see that this potential coincides, modulo redefinition of the coupling constants, with the image of the ``free'' system \p{ex3a}, despite the fact that
the prepotential \p{finpot1} is different from the one in \p{solPot2}. This difference plays an essential role after passing to the Hamiltonian with
spin variables, in which the corresponding potential acquires the form
\be\label{ho4}
 U= -\log(W) \quad \Rightarrow\quad \cV_U = J^{ab} J_{ab} \;\frac{ 2 \left( 24 m+M_1\right)^2 \epsilon^2 r^2 \left( 1 -\epsilon r^2\right)^2}
 {\left( 2 \epsilon^2 \left( m-\epsilon M_2\right) r^4 +\epsilon \left( 28 m + M_1 -4 \epsilon M_2\right) r^2 +2 \left(m -\epsilon M_2\right)\right)^2}.
\ee
Special cases of this potential are known. For example,
\bea
M_2 =\frac{ 32 m + M_1}{ 8 \epsilon} \qquad \Rightarrow &\quad&  \cV_U =  J^{ab} J_{ab} \; 8 \epsilon \left( V_{\textrm{Higgs}}-1 \right), \label{ho5} \\
M_2 = \frac{m}{\epsilon} \qquad \Rightarrow &\quad &  \cV_U =  J^{ab} J_{ab} \; 2 \frac{\left( 1-\epsilon r^2\right)^2}{r^2} \label{ho6}.
\eea
\setcounter{equation}{0}

\section{Conclusions}
We constructed $n$-dimensional ${\cal N}{=}\,4$ supersymmetric mechanics on arbitrary spaces with a metric $g_{ij}$. Besides reproducing the
curved WDVV equations already found in \cite{cWDVV} for the first prepotential, we obtained the curved version of the equations defining the second prepotential for the cases
of supercharges with and without spin variables.
For any solution of the curved WDVV equations on $so(n)$-invariant conformally flat spaces
we constructed admissible prepotentials on these spaces. A nice feature of our construction is the possibility to lift any ``flat'' prepotentials to isotropic spaces. Finally, we provided some interesting potentials for ${\cal N}{=}\,4$ mechanics on the (pseudo-)sphere.

A still unsolved task is a superspace description of our mechanics. To generalize
the superspace approach developed in \cite{AP1,AP2} clearly one will have to find new superspace irreducibility constraints for the $(1,4,3)$ supermultiplets which are covariant
under general coordinate transformations in the target space. A related task is to understand how the real target-space K\"{a}hler metric which
necessarily appears in the superspace approach \cite{AP1,AP2} is related with our component Hamiltonian description. These tasks will be considered elsewhere.

\section*{Acknowledgements}
We are grateful to M. Feigin and A. Veselov for stimulating discussions.
This work was partially supported by the Heisenberg-Landau program and has been performed within the ICTP Programs AF-04 and NT-04. 
The work of N.K. and S.K. was partially supported by Russian Science Foundation grant 14-11-00598, 
the one of A.S. by RFBR grants 15-02-06670 and 18-52-05002 Arm-a. The work of A.N. was partially 
supported by the Armenian State Committee of Science grants 15T-1C367 and 18RF-002. 
This article is based upon work from COST Action MP1405 QSPACE, supported by COST 
(European Cooperation in Science and Technology).

\end{document}